\documentclass[epj]{svjour}
\usepackage{graphics}
\usepackage{graphicx}
\usepackage{amsmath}
\usepackage{amsfonts}
\usepackage{subfigure}
\usepackage{float}
\usepackage{rotating}
\usepackage{afterpage}
\usepackage{widetext}
\newcommand{\phigg}{\phi_{\gamma\gamma}}
\begin{document}
\title{Deeply virtual Compton scattering using a positron beam in Hall-C at Jefferson Lab} 
\author{
A.~Afanasev\inst{9}
\and
I. Albayrak\inst{22}
\and
S.~Ali\inst{4}
\and
M.~Amaryan\inst{6}
\and
J.~R.~M.~Annand\inst{12}
\and
A.~Asaturyan\inst{5}
\and
V.~Bellini\inst{16}
\and
V.~V.~Berdnikov\inst{4}
\and
M.~Boer\inst{10}
\and
K.~Brinkmann\inst{14}
\and
W.~J.~Briscoe\inst{9}
\and
A.~Camsonne\inst{1}
\and
M.~Caudron\inst{2}
\and
L.~Causse\inst{2}
\and
M.~Carmignotto\inst{1}
\and
D.~Day\inst{13}
\and
M.~Defurne\inst{24}
\and
S.~Diehl\inst{14}
\and
R.~Ent\inst{1}
\and
P.~Chatagnon\inst{2}
\and
R.~Dupr\'e\inst{2}
\and
D.~Dutta\inst{17}
\and
M.~Ehrhart\inst{2}
\and
M.~A.~I.~Fernando\inst{23}
\and
T.~Forest\inst{11}
\and
M.~Guidal\inst{2}
\and
J.~Grames\inst{1}
\and
P.~Gueye\inst{15}
\and
S.~Habet\inst{2}
\and
D.~J.~Hamilton\inst{12}
\and
A.~Hobart\inst{2}
\and
T.~Horn\inst{4}
\and
C.~Hyde\inst{6}
\and
G.~Kalicy\inst{4}
\and
D.~Keller\inst{13}
\and
C.~Keppel\inst{1}
\and
M.~Kerver\inst{6}
\and
E.~Kinney\inst{18}
\and
H.-S.~Ko\inst{2}
\and
D.~Marchand\inst{2}
\and
P.~Markowitze\inst{8}
\and
M.~Mazouz\inst{3}
\and
M.~McCaughan\inst{1}
\and
B.~McKinnon\inst{12}
\and
A.~Mkrtchyan\inst{5}
\and
H.~Mkrtchyan\inst{5}
\and
M.~Muhoza\inst{4}
\and
C.~Mu\~noz~Camacho\inst{2}\thanks{Contact person: munoz@ijclab.in2p3.fr}%
\and
J.~Murphy\inst{7}
\and
P.~Nadel-Turonski\inst{19}
\and
S.~Niccolai\inst{2}
\and
G.~Niculescu\inst{20}
\and
R.~Novotny\inst{14}
\and
R.~Paremuzyan\inst{10}
\and
I.~Pegg\inst{4}
\and
K.~Price\inst{2}
\and
H.~Rashad\inst{6}
\and
J.~Roche\inst{7}
\and
R.~Rondon\inst{13}
\and
B.~Sawatzky\inst{1}
\and
V.~Sergeyeva\inst{2}
\and
S.~\v{S}irca\inst{21}
\and
A.~Somov\inst{1}
\and
I.~Strakovsky\inst{9}
\and
V.~Tadevosyan\inst{5}
\and
R.~Trotta\inst{4}
\and
H.~Voskanyan\inst{5}
\and
E.~Voutier\inst{2}
\and
B.~Wojtsekhowski\inst{1}
\and
S.~Wood\inst{1}
\and
S.~Zhamkochyan\inst{5}
\and
J.~Zhang\inst{13}
\and
S.~Zhao\inst{2}
\and
C.~Zorn\inst{1}%
}
%
%
\institute{
Thomas Jefferson National Accelerator Facility, 12000 Jefferson Avenue, Newport News, VA 23606, USA
\and
Universit\'e Paris-Saclay, CNRS/IN2P3, IJCLab Orsay, France
\and
Facult\'e des Sciences de Monastir, Tunisia
\and
The Catholic University of America, Washington, DC 20064, USA
\and
A.~Alikhanyan National Laboratory, Yerevan Physics Institute, Yerevan 375036, Armenia
\and
Old Dominion University, Norfolk, VA 23529, USA
\and
Ohio University, Athens, OH 45701, USA
\and
Florida International University, Miami, FL 33199, USA
\and
The George Washington University, Washington, DC 20052, USA
\and
University of New Hampshire, Durham, NH 03824, USA
\and
Idaho State University, Pocatello, ID 83209, USA
\and
University of Glasgow, Glasgow G12 8QQ, United Kingdom
\and
University of Virginia, Charlottesville, VA 22904, USA
\and
Universit\"at Gie\ss en Luwigstra\ss e 23, 35390 Gie\ss en, Deutschland
\and
Facility for Rare Isotope Beams, Michigan State University, 640 South Shaw Lane, East Lansing, MI 48824, USA
\and
Istituto Nazionale di Fisica Nucleare, Sezione di Catania, 95123 Catania, Italy
\and
Mississippi State University, MS 39762, USA
\and
University of Colorado, Boulder, CO 80309, USA
\and
Stony Brook University, Stony Brook, NY 11794, USA
\and
James Madison University, Harrisonburg, VA 22807, USA
\and
Faculty of Mathematics and Physics, University of Ljubljana, 1000 Ljubljana, Slovenia
\and
Akdeniz \"Universitesi, 07070 Konyaalti, Antalya, Turkey
\and
Hampton University Hampton, VA 23668, USA
\and
Commissariat \`a l'Energie Atomique, 91191 Gif-sur-Yvette, France
}

%
%
%
\date{Received: date / Revised version: date}
%
\abstract{
\vspace*{-0.2cm}We propose to use the High Momentum Spectrometer of Hall C combined with 
the Neutral Particle Spectrometer (NPS) to perform high precision 
measurements of the Deeply Virtual Compton Scattering (DVCS) cross 
section using a beam of positrons. The combination of measurements with 
oppositely charged incident beams is the only unambiguous way to 
disentangle the contribution of the DVCS$^2$ term in the photon 
electroproduction cross section from its interference with the 
Bethe-Heitler amplitude. This provides a stronger way to constrain the 
Generalized Parton Distributions of the nucleon. A wide range of 
kinematics accessible with an 11 GeV beam off an unpolarized proton 
target will be covered. The $Q^2-$dependence of each contribution will be measured independently.\vspace*{-0.2cm}
\PACS{
      {PACS-key}{discribing text of that key}   \and
      {PACS-key}{discribing text of that key}
     } 
} 
\maketitle

\section{Executive summary}
An exciting scientific frontier is the 3-dimensional exploration of nucleon (and nuclear) structure – nuclear femtography. Jefferson Lab with its high luminosity and expanded kinematic reach at 12-GeV will allow the detailed investigation of position and momentum distributions of partons inside protons and neutrons in the valence-quark region. The study of the Generalized Parton Distributions (GPDs) captures the images of the transverse position distributions of fast-moving quarks. The cleanest reaction to access GPDs is Deeply Virtual Compton Scattering (DVCS): 
$\gamma^*p\to\gamma p$.

A factorization theorem has been proven for DVCS in the Bjorken limit~\cite{Collins:1998be,Ji:1998xh}. It allows one to compute the DVCS 
amplitude as the product of some GPDs and a coefficient function that can be calculated perturbatively. GPDs are thus in very solid theoretical footing: at leading-twist level, all-order QCD-factorization theorems directly relate the GPDs to particular hard exclusive scattering processes. Therefore, GPDs are process-independent, universal quantities.

\begin{figure}[!bht]
    \centering
    \includegraphics[width=\linewidth]{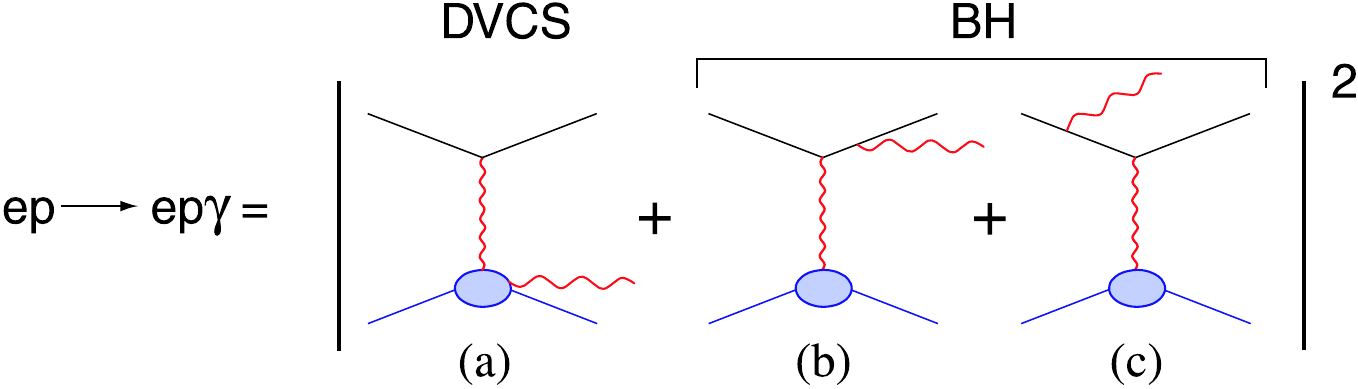}
    \caption{Illustration of the DVCS (a) and Bethe-Heitler (b and c) processes.}
    \label{fig:bbhh}
\end{figure}

DVCS interferes with the so-called Bethe-Heitler (BH) process, where the lepton scatters elastically off the nucleon and emits a high energy photon before or after the interaction (see Fig.~\ref{fig:bbhh}). 
The BH amplitude $\mathcal T^{BH}$ is electron charge even. On the other hand, the DVCS amplitude $\mathcal T^{DVCS}$ is electric charge odd, i.e. its contribution has different sign for electron vs. positron scattering. DVCS and BH are indistinguishable and the photon electroproduction amplitude squared that we can measure is therefore decomposed as:
\begin{equation}
    |\mathcal T(\pm ep\to \pm ep\gamma)|^2=|\mathcal T^{BH}|^2+|\mathcal T^{DVCS}|^2\mp\mathcal I\,,
\label{eq:bhint}
\end{equation}
where the $\pm$ signs correspond to the charge of the incident beam. The $\mathcal T^{BH}$ amplitude is written in terms of the nucleon form factors, and is real at the leading order in QED.
The $|\mathcal T^{DVCS}|^2$ contribution is closest to a direct Compton scattering cross section and as such gives direct information on nucleon structure – it depends on bilinear combinations of GPDs.

Equation~\ref{eq:bhint} shows how combining DVCS measurements with electrons and positrons not only can cleanly isolate the $|\mathcal T^{DVCS}|^2$ term but also the interference term $\mathcal I$. This interference term gives direct linear access to DVCS at the amplitude level, thanks to its interference with the known Bethe-Heitler amplitude. Similar as in spin-dependent scattering, such interferences can lead to extremely rich angular structure: $\mathcal I = 2\mathcal T^{BH}\mathcal Re(\mathcal T^{DVCS})$. 

{ The availability of positron beams} thus can lead to direct access to nucleon structure carried in the DVCS amplitude, and in addition a cleaner access to the $|\mathcal T^{DVCS}|^2$ term.


\section{Introduction}
Deeply Virtual Compton Scattering (DVCS) refers to  the reaction
$\gamma^*(q)P(p)\rightarrow P(p')\gamma(q')$ in the Bjorken limit of Deep Inelastic
Scattering (DIS). Experimentally, we can access DVCS through electroproduction of real photons $e(k)P(p)\to e(k')P(p')\gamma(q')$, where the DVCS amplitude interferes with the so-called Bethe-Heitler (BH) process. The BH contribution is calculable in QED since it corresponds to the emission of the photon by the incoming or the outgoing electron. 

DVCS is the simplest probe of a new class
of light-cone (quark) matrix elements, called 
Generalized Parton Distributions (GPDs)~\cite{Ji:1996ek}.  The GPDs offer the
exciting possibility of the first ever spatial images of the
quark waves inside the proton, as a function of their wavelength~\cite{Mueller:1998fv,Ji:1996nm,Ji:1996ek,Ji:1997gm,Radyushkin:1997ki,Radyushkin:1996nd}.
The correlation of transverse spatial and longitudinal momentum
information contained in the GPDs provides a new tool
to evaluate the contribution of quark orbital angular momentum
to the proton spin.

GPDs enter the DVCS cross section through integrals over the quark momentum fraction $x$, called Compton Form Factors (CFFs).
CFFs are defined in terms of the vector GPDs
$H$ and $E$, and the axial vector GPDs $\widetilde{H}$ and $\widetilde{E}$~\cite{Ji:1996ek}.
For example ($f\in\{u,d,s\}$) \cite{Belitsky:2001ns}:
\begin{multline}
{\mathcal H}(\xi,t) = \sum_{f}  \left[\frac{e_f}{e}\right]^2
\Biggl\{
         i\pi     \left[H_f(\xi,\xi,t) - H_f(-\xi,\xi,t)\right]
  \\
   +
 {\mathcal P}
    \int_{-1}^{+1} dx
\left[ \frac{1}{\xi-x} - \frac{1}{\xi+x} \right]   H_f(x,\xi,t)\Biggr\},
\label{eq:CFF}
\end{multline}
where $t=(p-p')^2$ is the momentum transfer to the nucleon and skewness variable $\xi$ is defined as $\xi=-\overline{q}^2/(\overline{q}\cdot \overline{p})\approx x_{\rm B}/(2-x_{\rm
B})$, with $\overline{q}= (q+q')/2$ and $\overline{p}=p+p'$.

Thus, the imaginary part accesses GPDs along the line $x=\pm\xi$,
whereas the real part probes GPD integrals over $x$. The `diagonal' GPD, $H(\xi,\xi,t=\Delta^2)$ is not a positive-definite probability
density, however it is a transition density with the momentum transfer $\Delta_\perp$ Fourier-conjugate to the transverse distance $r$
between the active parton and the center-of-momentum of the spectator partons in the target \cite{Burkardt:2007sc}.
  Furthermore, the real part of
the Compton Form Factor is determined by a dispersion integral over the diagonal $x=\pm \xi$ plus the $D$-term \cite{Teryaev:2005uj,Anikin:2007yh,Anikin:2007tx,Diehl:2007ru}:
\begin{widetext}
\begin{equation}
  \Re\text{e}\left[\mathcal H (\xi,t)\right] = \int_{-1}^1 dx \left\{\left[H(x,x,t)+H(-x,x,t)\right]
  \left[\frac{1}{\xi-x} - \frac{1}{\xi+x}\right] 
  + 2 \frac{D(x,t)}{1-x} \right\}
\end{equation}
\end{widetext}
The $D$-term~\cite{Polyakov:1999gs} only has support in the region $|x|<\xi$ in which the GPD is determined by $q\overline{q}$ exchange in the $t$-channel.


\section{Physics goals}

In this experiment we propose to exploit the charge dependence provided by the use of a positron beam in order to cleanly separate the DVCS$^2$ term from the DVCS-BH interference in the photon electroproduction cross section.

\begin{figure}[b]
\begin{center}
\includegraphics[width=\linewidth]{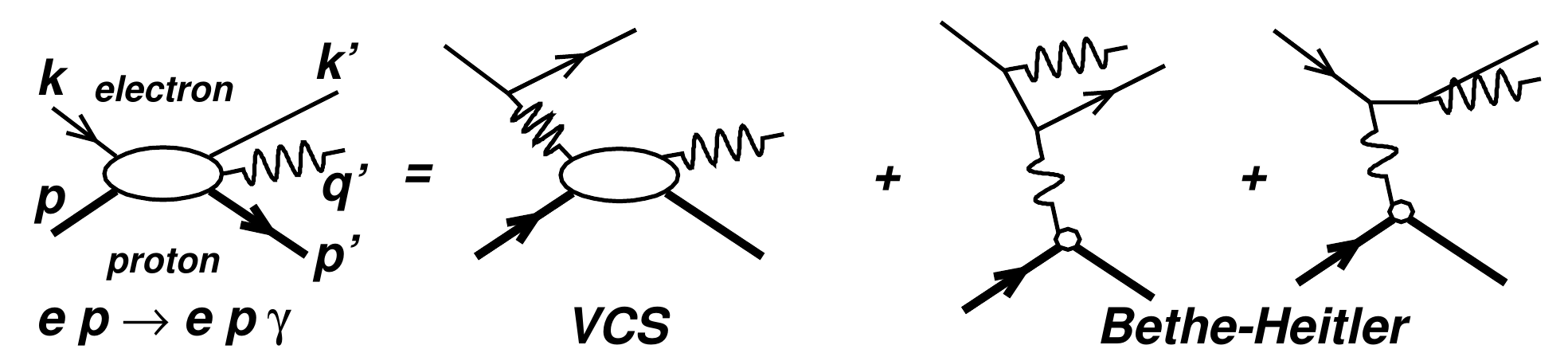}
\caption{\baselineskip 13 pt
Lowest order QED amplitude for the $ep\rightarrow ep\gamma$ reaction.
 The momentum four-vectors of all external particles are labeled at left.
The net four-momentum transfer to the proton is 
$\Delta_\mu=(q-q')_\mu=(p'-p)_\mu$. In the virtual Compton scattering
(VCS) amplitude, the (spacelike) virtuality of the incident photon is
$Q^2=-q^2=-(k-k')^2$.  In the Bethe-Heitler (BH) amplitude, the virtuality
of the incident photon is $-\Delta^2=-t$.  Standard $(e,e')$ invariants
are $s_e=(k+p)^2$, $x_B=Q^2/(2q\cdot p)$ and $W^2=(q+p)^2$.}
\label{fig:epepg}
\end{center}
\end{figure}

 The photon electroproduction cross section of a polarized lepton beam of energy $E_b$  off
 an unpolarized target of mass $M$ is sensitive to the coherent interference of the DVCS amplitude
 with the Bethe-Heitler amplitude (see Fig. \ref{fig:epepg}). It was derived in \cite{Ji:1996nm} and can be written as:
\begin{multline}
\frac{d^5\sigma(\lambda,\pm e)}{d^5\Phi} =
\frac{d\sigma_0}{dQ^2 dx_B}
\left| \mathcal T^{BH}(\lambda) \pm \mathcal T^{DVCS}(\lambda)
     \right|^2/|e|^6 \nonumber \\
 =  \frac{d\sigma_0}{dQ^2 dx_B}   \left[
       \left|\mathcal T^{BH}(\lambda) \right|^2 +
        \left| \mathcal T^{DVCS}(\lambda)\right|^2  \mp
        \mathcal I(\lambda)   \right]\frac{1}{e^6} \label{eq:dsigDVCS}
\end{multline}
 \begin{eqnarray}
   \frac{d\sigma_0}{dQ^2 dx_B} &=&
\frac{\alpha_{\rm QED}^3}{16\pi^2(s_e-M^2)^2 x_B} 
\frac{1}{\sqrt{1+\epsilon^2}}   \\
\epsilon^2 &=& 4 M^2 x_B^2/Q^2 \nonumber \\
s_e&=& 2 M E_b + M^2 \nonumber 
\label{eq:dsig0}
\end{eqnarray}
where $d^5\Phi=dQ^2 dx_B d\phi_e dt d\phi_{\gamma\gamma}$, 
$\lambda$ is the electron helicity and the $+$($-$) stands for the sign of the charge of the lepton beam.  
The BH contribution is calculable in QED, given our $\approx 1\%$  knowledge of the
proton elastic form factors at small momentum transfer. The other two
contributions to the cross section, the interference and the DVCS$^2$ terms,
provide complementary information on GPDs. It is possible to exploit the structure of the
cross section as a function of the angle $\phi_{\gamma\gamma}$ between the leptonic and hadronic plane to separate
up to a certain degree the different contributions to the total cross section~\cite{Diehl:1997bu}. The  angular
separation can be supplemented by an beam energy separation. The energy separation has been successfully used in previous experiments~\cite{Defurne:2017paw} at 6 GeV and is the goal of already approved experiment at 12 GeV~\cite{E12-13-010}.

The $|\mathcal T^{BH}|^2$ term is given in \cite{Belitsky:2001ns},
Eq.~(25), and only its general form is reproduced here:
\begin{multline}
 |\mathcal T^{BH}|^2 =
\frac{e^6}{x_B^2 t y^2 (1+\epsilon^2)^2  \mathcal P_1(\phigg) \mathcal P_2(\phigg)}\\
\sum_{n=0}^2 c_n^{BH} \cos(n\phi_{\gamma\gamma})\,.
\end{multline}

The harmonic terms $c_n^{BH}$ depend upon
 bilinear combinations of the ordinary elastic form factors
$F_1(t)$ and $F_2(t)$ of the proton.  The factors $\mathcal P_i$ are the electron
propagators in the BH amplitude~\cite{Belitsky:2001ns}.

 The interference term in Eq.~(\ref{eq:dsigDVCS}) is a
linear combination of GPDs, whereas the DVCS$^2$ term is a
bilinear combination of GPDs.  These terms have the following harmonic structure:
\begin{widetext}
\begin{eqnarray}
\mathcal I &=&
\frac{e^6}{x_B y^3 \mathcal P_1(\phigg) \mathcal P_2(\phigg) t }
\left\{ c_0^{\mathcal I} + \sum_{n=1}^3
     \left[  
   c_n^{\mathcal I}\cos(n\phigg)
        +    \lambda s_n^{\mathcal I}\sin(n\phigg) \right] \right\}
\label{eq:IntPhi}
\end{eqnarray}
\begin{eqnarray}
\left| \mathcal T^{DVCS}(\lambda) \right|^2  &=&
\frac{e^6}{y^2 Q^2} \left\{
c_0^{DVCS} + \sum_{n=1}^2 
  \left[   c_n^{DVCS} \cos(n\phigg) - \lambda s_1^{DVCS} \sin(\phigg) \right]\right\}
\label{eq:DVCSPhi}
\end{eqnarray}
\end{widetext}

The $c_0^{DVCS, \mathcal I}$, and $(c,s)_1^{\mathcal I}$ harmonics are dominated by twist-two  GPD terms, although they do have twist-three admixtures that must be quantified by
the $Q^2$-dependence of each harmonic.  The $(c,s)_1^{DVCS}$ and $(c,s)_2^{\mathcal I}$ harmonics are dominated by twist-three matrix elements, although the same twist-two GPD terms
also contribute (but with smaller kinematic coefficients than in the lower Fourier terms).  The $(c,s)_2^{DVCS}$ and $(c,s)_3^{\mathcal I}$ harmonics
stem from  twist-two double helicity-flip gluonic GPDs alone. They are formally suppressed by $\alpha_s$ and will be neglected here. They do not mix, however, with the  twist-two quark amplitudes.
The exact expressions of these harmonics in terms of the quark Compton Form Factors (CFFs) of the nucleon are given in~\cite{Belitsky:2010jw}.

Equation~(\ref{eq:dsigDVCS}) shows how a positron beam, together with measurements with electrons, provides a way to separate without any assumptions the DVCS$^2$ and BH-DVCS interference contributions to the cross section. With electrons alone, the only approach to this separation is to use the different beam energy dependence of the DVCS$^2$ and BH-DVCS interference. This is the strategy that will be used in approved experiment E12-13-010. However, as recent results have shown~\cite{Defurne:2017paw} this technique has limitations due to the need to include power corrections to fully describe the precise azimuthal dependence of the DVCS cross sections.

A positron beam, on the other hand, will be able to pin down each individual term. The $Q^2-$dependence of each of them can later be 
used to study the nature of the higher twist contributions by comparing it to the predictions of the leading twist diagram.

A positron beam can also be used to measure the corresponding beam charge asymmetry defined as:
\begin{equation}
    A_C(\phi_{\gamma\gamma})=\frac{d\sigma^+(\phi_{\gamma\gamma})-d\sigma^-(\phi_{\gamma\gamma})}{d\sigma^+(\phi_{\gamma\gamma})+d\sigma^-(\phi_{\gamma\gamma})}\,,
    \label{eq:bca}
\end{equation}
which is easier experimentally. This measurement was pioneered by the HERMES collaboration~\cite{Airapetian:2006zr}. A drawback, however, is that it depends non-linearly on the DVCS amplitudes because of the denominator.
One can further project the beam charge asymmetry on the various harmonics:
\begin{equation}
A_C^{\cos{(n\phi)}}=\frac{2-\delta_{n0}}{2\pi}\int^\pi_{-\pi} d\phi_{\gamma\gamma}\cos{(n\phi_{\gamma\gamma})}A_C(\phi_{\gamma\gamma})\,,
\end{equation}
The $A_C^{\cos{(n\phi)}}$ is governed by the $c^{\mathcal I}_{n}$ of Eq.~(\ref{eq:IntPhi}). Nonetheless, because of the
$\phi_{\gamma\gamma}$-dependent denominator in \eqref{eq:bca}, it is contaminated by all other harmonics as well~\cite{Braun:2014sta}. Absolute cross-section measurements are thus needed to cleanly measure the interference term without any contamination.

GPDs appear in the DVCS cross section under convolution integrals, usually called Compton Form Factors (CFFs): $\mathcal{F}_{\mu \nu}$, where $\mu$ and $\nu$ are the helicity states of the virtual photon and the outgoing real photon, respectively.
The interference between BH and DVCS provides a way to independently access the real and imaginary parts of CFFs. At leading-order, the imaginary part of the helicity-conserving $\mathcal{F}_{++}$ is directly related to the corresponding GPD at $x=\xi$:
\begin{eqnarray}
\mathcal R\text{e}\,\mathcal{F}_{++}&=& \mathcal{P} \int_{-1}^{1}dx\left[\frac{1}{x-\xi} -\kappa \frac{1}{x+\xi}\right]F(x,\xi,t)\;,\nonumber\\
\mathcal I\text{m}\,\mathcal{F}_{++}&=& -\pi\left[F(\xi,\xi,t)+\kappa F(-\xi,\xi,t)\right]\;,
\label{eq:dispersion}
\end{eqnarray}
where $\kappa=-1$ if $F \in\{H,E\}$ and $1$ if $F \in\{\widetilde{H},\widetilde{E}\}$. Recent phenomenology uses the leading-twist (LT) and leading-order (LO) approximation in order to extract or parametrize GPDs, which translates into neglecting $\mathcal{F}_{0+}$ and $\mathcal{F}_{-+}$ and using the relations of Eq.~\ref{eq:dispersion}~\cite{Kumericki:2009uq,Kumericki:2016ehc,Dupre:2016mai}.

The scattering amplitude is a Lorentz invariant quantity, but the deeply virtual scattering process nonetheless defines a preferred axis (light-cone axis) for describing the scattering process. At finite $Q^2$ and non-zero $t$, there is an ambiguity in defining this axis, though all definitions converge as
$Q^2\to \infty$ at fixed $t$. Belitsky et al.~\cite{Belitsky:2012ch} decompose the DVCS amplitude in terms of photon-helicity states where the light-cone axis is defined in the plane of the four-vectors $q$ and $P$.  This leads to the CFFs defined previously. Recently, Braun \emph{et al.}~\cite{Braun:2014sta} proposed an alternative decomposition which defines the light cone axis in the plane formed by $q$ and $q'$ and argue that this is  more convenient to account for kinematical power corrections of $\mathcal{O}(t/Q^2)$ and $\mathcal{O}(M^2/Q^2)$. The bulk of these corrections can be included by rewriting the CFFs $\mathcal{F}_{\mu \nu}$ in terms of $\mathbb{F}_{\mu \nu}$ using the following map~\cite{Braun:2014sta}:
\begin{eqnarray}
\label{eq:BMPtoBMJ}
\mathcal{F}_{++}=&\mathbb{F}_{++}+\frac{\chi}{2}\left[\mathbb{F}_{++}+ \mathbb{F}_{-+}\right]-\chi_0 \mathbb{F}_{0+}\;,\\
\mathcal{F}_{-+}=&\mathbb{F}_{-+}+\frac{\chi}{2}\left[\mathbb{F}_{++}+ \mathbb{F}_{-+}\right]-\chi_0 \mathbb{F}_{0+}\;,\label{eq:BMPtoBMJ1}\\
\mathcal{F}_{0+}=&-(1+\chi)\mathbb{F}_{0+}+\chi_0\left[\mathbb{F}_{++}+ \mathbb{F}_{-+}\right]\;,\label{eq:BMPtoBMJ2}
\end{eqnarray}
where kinematic parameters $\chi_0$ and $\chi$ are defined  as follows (Eq.~48 of Ref~\cite{Braun:2014sta}):
\begin{eqnarray}
\chi_0=&\displaystyle\frac{\sqrt{2}Q\widetilde{K}}{\sqrt{1+\epsilon^2}(Q^2+t)}&\propto \frac{\sqrt{t_{min}-t}}{Q}\;,\\
\chi=&\displaystyle\frac{Q^2-t+2x_Bt}{\sqrt{1+\epsilon^2}(Q^2+t)}-1&\propto \frac{t_{min}-t}{Q^2}\;.
\end{eqnarray}

Within the $\mathbb{F}_{\mu \nu}$-parametrization, the leading-twist and leading-order approximation consists in keeping $\mathbb{F}_{++}$ and neglecting both $\mathbb{F}_{0+}$ and  $\mathbb{F}_{-+}$. Nevertheless, as a consequence of Eq.~(\ref{eq:BMPtoBMJ1}) and (\ref{eq:BMPtoBMJ2}), $\mathcal{F}_{0+}$ and $\mathcal{F}_{-+}$ are no longer equal to zero since proportional to $\mathbb{F}_{++}$. The functions that can be extracted from data to describe the three dimensional structure of the nucleon become:
\begin{equation}
\mathcal{F}_{++}=(1+\frac{\chi}{2})\mathbb{F}_{++},\,
\mathcal{F}_{0+}=\chi_0\mathbb{F}_{++},\,
\mathcal{F}_{-+}=\frac{\chi}{2}\mathbb{F}_{++}.
\end{equation}
A numerical application gives $\chi_0=$0.25 and $\chi=$0.06 for $Q^2$=2~GeV$^2$, x$_B$=0.36 and $t=-0.24$~GeV$^2$. Considering the large size of the parameters $\chi_0$ and $\chi$, these kinematical power corrections cannot be neglected in precision DVCS phenomenology, in particular in order to unambiguously extract the CFFs. Indeed, when the beam energy changes, not only do the contributions of the DVCS-BH interference and DVCS$^2$ terms change but also the polarization of the virtual photon changes, 
thereby modifying the weight of the different helicity amplitudes.

The calculation of power corrections to DVCS is one of the most important theory advances in DVCS in recent years. BMP~\cite{Braun:2014sta} have convincingly shown that in JLab kinematics target mass corrections can be sizeable and cannot be neglected.

\section{Experimental setup}
We propose to make a precision coincidence setup measuring charged particles (scattered positrons) with the existing HMS and photons using the Neutral Particle Spectrometer (NPS), currently under construction. The NPS facility consists of a PbWO$_4$ crystal calorimeter and a sweeping magnet in order to reduce electromagnetic backgrounds. A high luminosity spectrometer and calorimeter (HMS+PbWO$_4$) combination proposed in Hall C is ideally suited for such measurements.

The sweeping magnet will allow to achieve low-angle photon detection. Detailed background simulations show that this setup allows for $\ge 10 \mu A$ beam current on a 10\,cm long cryogenic LH2 target at the very smallest NPS angles, and much
higher luminosities at larger $\gamma,\pi^0$ angles~\cite{E12-13-010}.

\subsection{High Momentum Spectrometer}
 The magnetic spectrometers benefit from relatively small point-to-point uncertainties, which are crucial for absolute cross section measurements. In particular, the optics properties and the acceptance of the HMS have been studied extensively and are well understood in the kinematic e between 0.5 and 5 GeV, as evidenced by more than 200 L/T separations ($\sim$1000 kinematics)~\cite{Liang:2004tj}. The position of the elastic peak has been shown to be stable to better than 1 MeV, and the precision rail system and rigid pivot connection have provided reproducible spectrometer pointing for about a decade.

\subsection{Photon detection: the neutral particle spectrometer (NPS)}

\begin{figure}[!htbp]
\centering
\subfigure[\label{fig:exp-setup} \it ]
{
\includegraphics[bb=6 20 1508 958, clip,width=\linewidth]{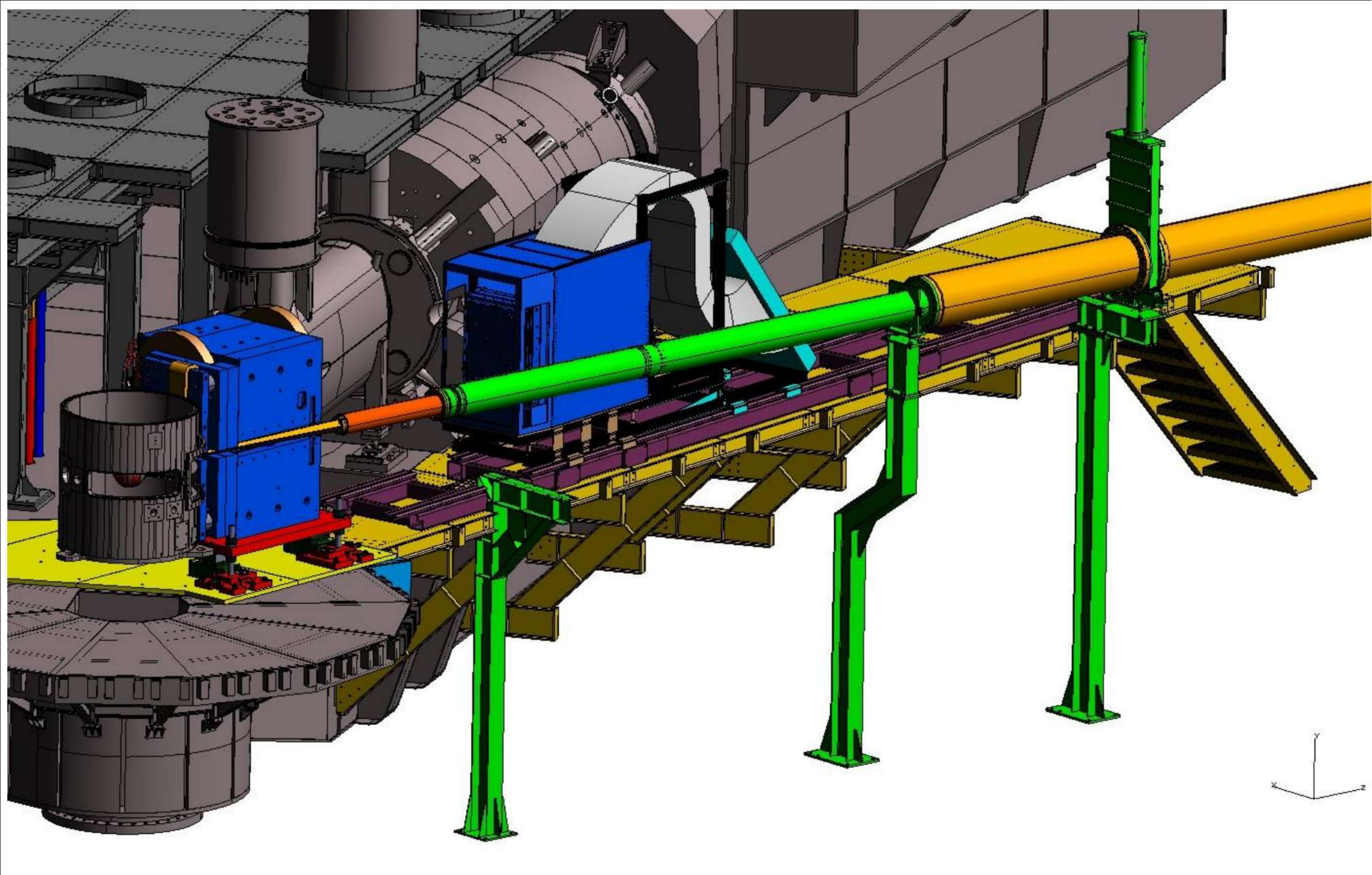}}
\hspace{1cm}
\subfigure[\label{fig:PbWO4_block} \it ] 
{
\includegraphics[width=2.9in]{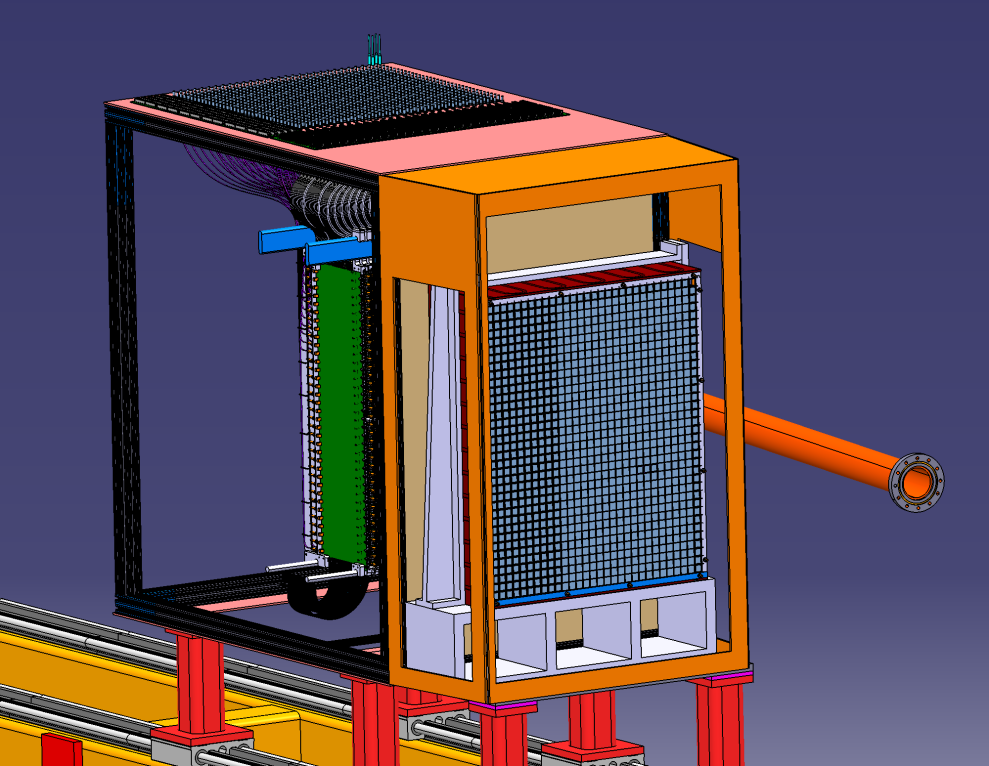}}
\hspace{1cm}
\subfigure[\label{fig:PbWO4_block2} \it ] 
{\includegraphics[width=2.9in]{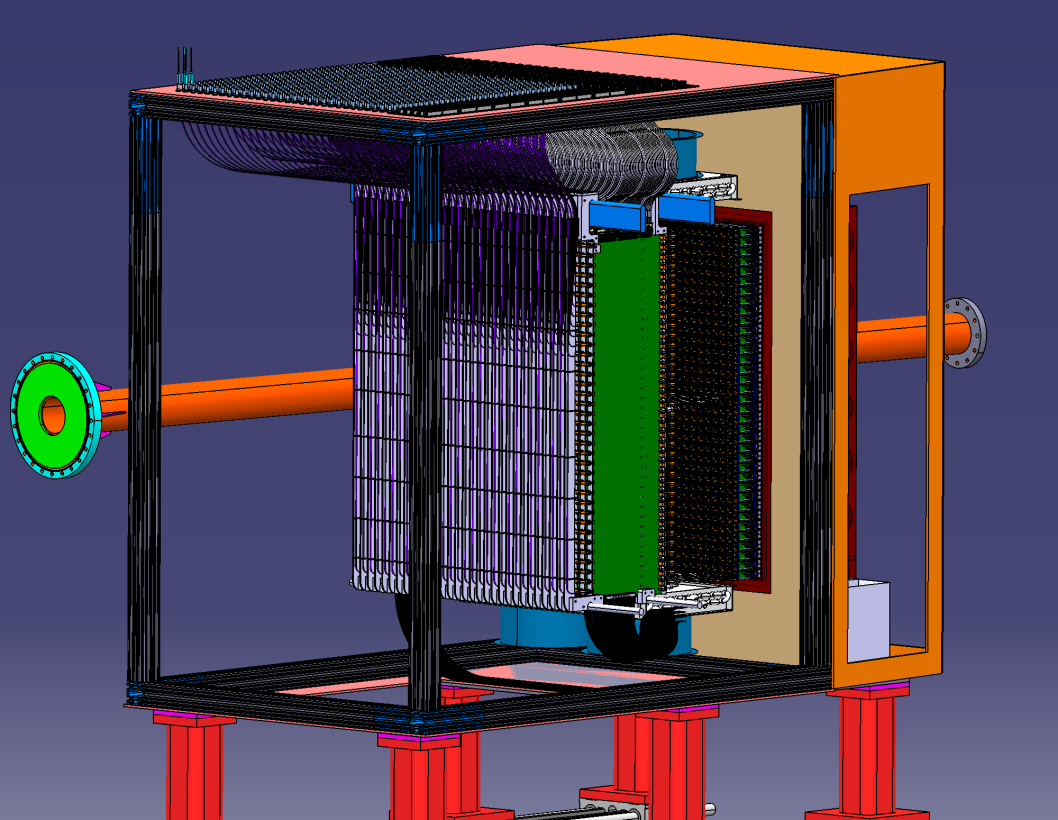}}
\caption{\label{fig:pi0_detector_hallc} \baselineskip 13 pt
 (a)~The DVCS detector in Hall C. The cylinder in the left is the (1 m diameter) vacuum chamber containing the 10-cm long liquid-hydrogen target. The NPS sweeping magnet and calorimeter are standing on the yellow platform of the SHMS, which will be used as carriage to support them. The HMS (not shown) placed on the other side of the beam line will be used to detect the scattered positrons.
(b) Front view of the NPS calorimeter showing the PbWO$_4$ crystal array. (c) Back view of the calorimeter showing the PMT voltage dividers and the vertical PCB distribution boards which bring HV and transfer the PMT signal to the read-out electronics.}
\end{figure}

We will use the general-purpose and remotely rotatable NPS system for Hall C. A layout of NPS standing in the SHMS carriage is shown in Fig.~\ref{fig:exp-setup}. The NPS system consists of the following elements:

\begin{itemize}
\item{A sweeping magnet providing 0.3 Tm field strength. }
\item{A neutral particle detector consisting of 1080 PbWO$_4$ crystals in a temperature controlled frame, comprising a 25 msr device at a distance of 4 meters.}
\item{Essentially deadtime-less digitizing electronics to independently sample 
the entire pulse form for each crystal allowing for background subtraction and 
identification of pile-up in each signal.}
\item{A new set of high-voltage distribution bases with built-in amplifiers 
for operation in high-rate environments.}
\item{Cantilevered platforms on the SHMS carriage, to allow for precise and 
remote rotation around the Hall C pivot of the full photon detection 
system, over an angle range between 6 and 30 degrees.}
\item{A dedicated beam pipe with as large critical angle as possible to reduce
backgrounds beyond the sweeping magnet.}
\end{itemize}

\subsubsection*{The PbWO$_4$ electromagnetic calorimeter}
The energy resolution of the photon detection is the limiting factor of the experiment. Exclusivity of the reaction is ensured by the missing mass technique (see section~\ref{sec:mm2}) and the missing-mass resolution is dominated by the energy resolution of the calorimeter.

We plan to use a PbWO$_4$ calorimeter 56 cm wide and 68 cm high. This corresponds to 28 by 34 PbWO$_4$ crystals of 2.05 by 2.05 cm$^2$ (each 20.0 cm long).
We have added one crystal on each side to properly capture showers, and thus 
designed our PbWO$_4$ calorimeter to consist of 30 by 36 PbWO$_4$ crystals, or 
60 by 72 cm$^2$. This amounts to a requirement of 1080 PbWO$_4$ crystals.

To reject very low-energy background, a thin absorber could be installed in 
front of the PbWO$_4$ detector. The space between the sweeper magnet and the proximity of the PbWO$_4$ detector will be enclosed within a vacuum channel (with a thin exit window, further reducing low-energy background) to minimize the decay photon conversion in air.

Given the temperature sensitivity of the scintillation light output of the
PbWO$_4$ crystals, the entire calorimeter must be kept at a constant 
temperature, to within $0.1^\circ$ to guarantee 0.5\% energy stability for 
absolute calibration and resolution. The high-voltage dividers on the PMTs may 
dissipate up to several hundred Watts, and this power similarly must not 
create temperature gradients or instabilities in the calorimeter. 
The calorimeter will thus be thermally isolated and be surrounded on 
all four sides by water-cooled copper plates.

At the anticipated background rates, 
pile-up and the associated baseline shifts can adversely affect the 
calorimeter resolution, thereby constituting the limiting factor for the beam 
current. The solution is to read out a sampled signal, and perform offline 
shape analysis using a flash ADC (fADC) system. 
New HV distribution bases with built-in pre-amplifiers will allow for operating the PMTs at lower 
voltage and lower anode currents, and thus protect the photocathodes or 
dynodes from damage.

The PbWO$_4$ crystals are 2.05 x 2.05 cm$^2$. 
The typical position resolution is 2-3 mm. Each crystal covers 5 mrad, and the 
expected angular resolution is 0.5-0.75 mrad, which is comparable with the 
resolutions of the HMS and SOS, routinely used for Rosenbluth separations in 
Hall C. 

To take full advantage of the high-resolution crystals while operating in a high-background environment,
modern flash ADCs will be used to digitize the signal. They continuously sample the signal every 4\,ns,
storing the information in an internal FPGA memory. When a trigger is received, the samples in a programmable
window around the threshold crossing are read out for each crystal that fired. Since the readout of the FPGA
does not interfere with the digitizations, the process is essentially deadtime free.

\subsection{Exclusivity of the DVCS reaction}
\label{sec:mm2}
\begin{figure}[h!]
\begin{center}
\includegraphics[width=0.59\linewidth]{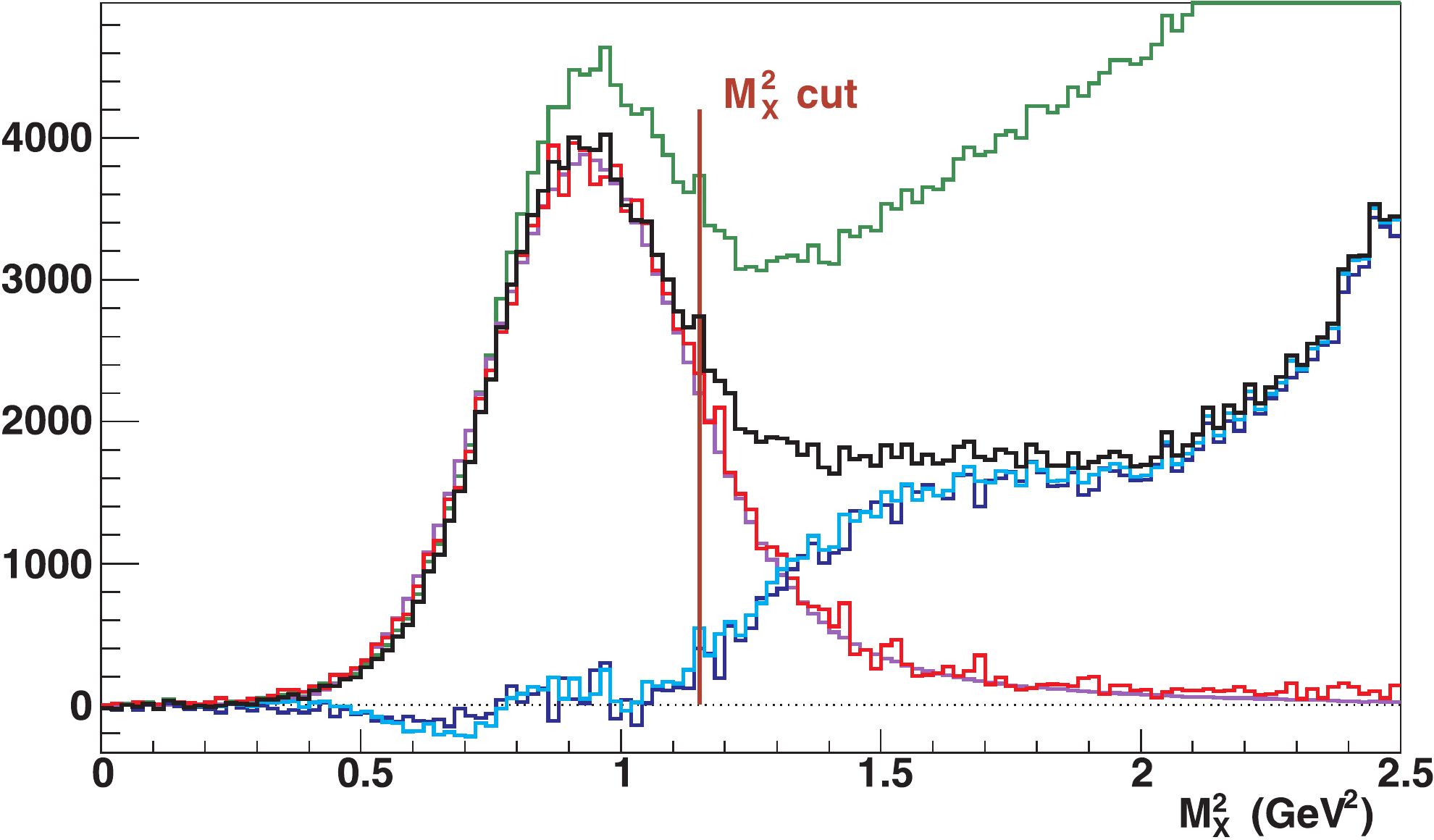}
\includegraphics[width=0.39\linewidth]{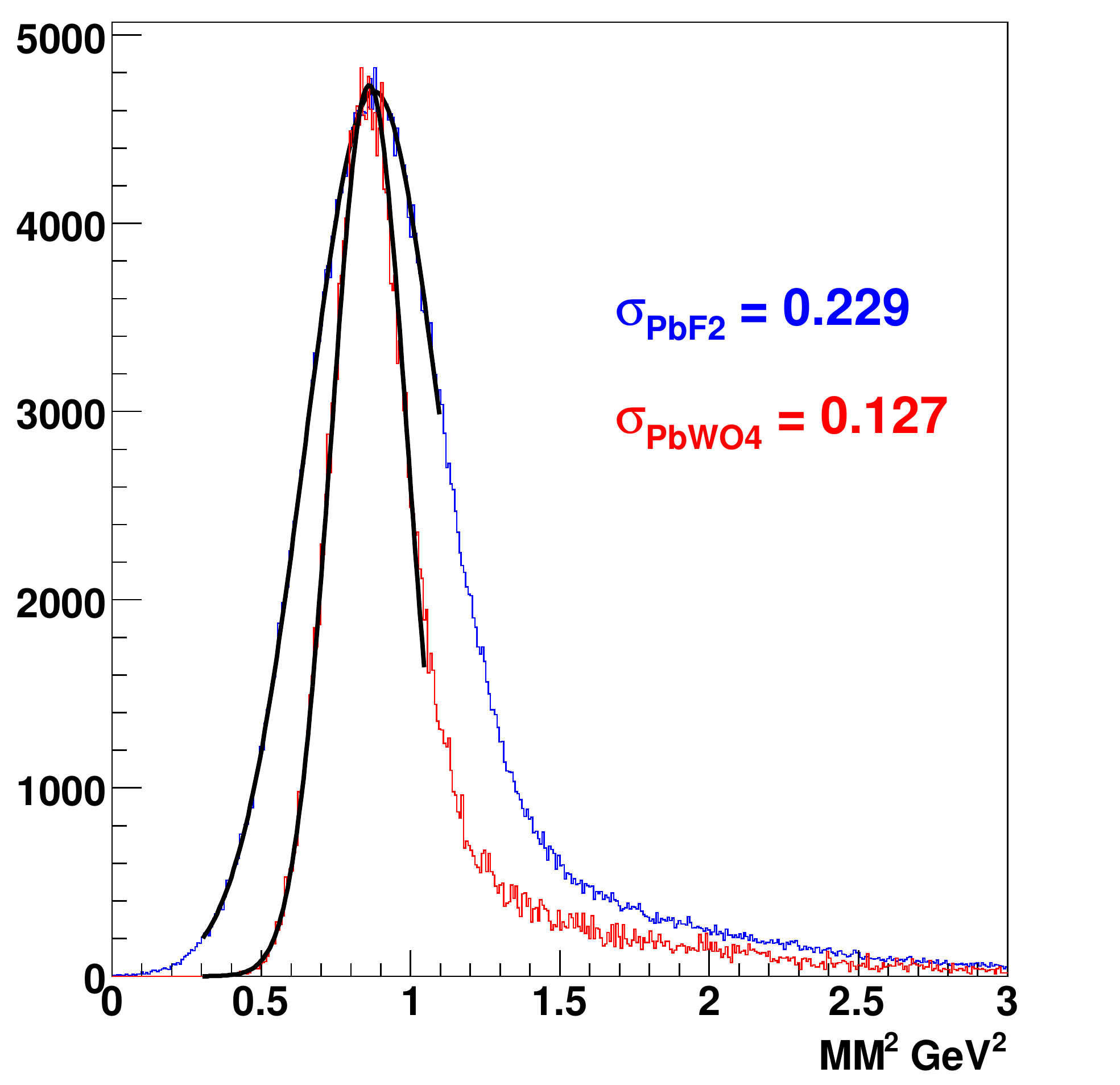}
\caption{\baselineskip 13 pt
Left: Missing mass squared in E00-110 for H$(e,e'\gamma)X$ events (green curve) at
$Q^2=2.3$ GeV$^2$ and $-t\in[0.12,0.4]$ GeV$^2$, integrated over the
azimuthal angle of the photon $\phi_{\gamma\gamma}$. The black curve shows
the data once the H$(e,e'\gamma)\gamma X'$ events have been subtracted.
The other curves are described in the text. Right: Projected missing mass resolution for a similar kinematic setting ($E_b=6.6$~GeV, $Q^2=3$~GeV$^2$, $x_B=0.36$). By using PbWO$_4$ instead of PbF$_2$, the missing mass resolution will be considerably improved. Values are given in Tab.~\ref{tab:DVCS-Kin} and are to be compared to the the value $\sigma(M_X^2)=0.2$~GeV$^2$ obtained in previous experiments in Hall A and showed in this figure (left).
}
\label{fig:mm2}
\end{center}
\end{figure}

The exclusivity of the DVCS reaction will be based on the missing mass
technique, successfully used during Hall A experiments E00-110 and E07-007 with a PbF$_2$ calorimeter. 
Fig.~\ref{fig:mm2} presents the missing mass squared obtained in E00-110 for
H$(e,e'\gamma)X$ events, with coincident electron-photon detection.
After
subtraction of an accidental coincidence sample, our data is essentially
background free: we have negligible contamination of non-electromagnetic
events in the HRS and PbF$_2$ spectra.  However, in addition to H$(e,e'\gamma) p$,
we do have the following competing channels:   H$(e,e'\gamma) p\gamma$ from $e p \rightarrow e
\pi^0 p$, $e p \rightarrow e \pi^0 N\pi$, $e p \rightarrow e \gamma N\pi$,
$e p \rightarrow e \gamma N\pi\pi\ldots$. From symmetric (lab-frame)
$\pi^0$-decay, we obtain a high statistics sample of H$(e,e'\pi^0)X'$
events, with two photon clusters in the PbF$_2$ calorimeter. From these
events, we determine the statistical sample of [asymmetric]
H$(e,e'\gamma)\gamma X'$ events that must be present in our
H$(e,e'\gamma)X$ data. The $M_X^2$ spectrum displayed in black in
Fig.~\ref{fig:mm2} was obtained after subtracting this $\pi^0$ yield from
the total (green) distribution. This is a $14\%$ average subtraction in
the exclusive window defined by '$M_X^2$\,cut' in Fig.~\ref{fig:mm2}.
Depending on the bin in $\phi_{\gamma\gamma}$ and $t$, this subtraction
varies from 6\% to 29\%. After our $\pi^0$
subtraction, the only remaining channels, of type H$(e,e'\gamma)N\pi$,
$N\pi\pi$, {\it etc.\/} are kinematically constrained to $M_X^2 >
(M+m_\pi)^2$. This is the value ('$M_X^2$\,cut' in Fig.~\ref{fig:mm2}) we
chose for truncating our integration. Resolution effects can cause the
inclusive channels to contribute below this cut. To evaluate this possible
contamination, during E00-110 we used an additional proton array (PA) of 100 plastic
scintillators. The PA subtended a solid angle (relative to the nominal
direction of the {\bf q}-vector) of $18^\circ<\theta_{\gamma p}<38^\circ$
and $45^\circ < \phi_{\gamma p} = 180^\circ-\phi_{\gamma\gamma} <
315^\circ$, arranged in 5 rings of 20 detectors. For H$(e,e'\gamma)X$
events near the exclusive region, we can predict which block in the PA
should have a signal from a proton from an exclusive H$(e,e'\gamma p)$
event.  The red histogram is the $X=(p+y)$ missing mass squared
distribution for H$(e,e'\gamma p)y$ events in the predicted PA block, with
a signal above an effective threshold $30$ MeV (electron equivalent). The
blue curve shows our inclusive yield, obtained by subtracting the
normalized triple coincidence yield from the H$(e,e'\gamma)X$ yield. The
(smooth)  violet curve shows our simulated H$(e,e'\gamma)p$ spectrum,
including radiative and resolution effects, normalized to fit the data for
$M_X^2\le M^2$. The cyan curve is the estimated inclusive yield obtained
by subtracting the simulation from the data. The blue and cyan curves are
in good agreement, and show that our exclusive yield has less than $2\%$
contamination from inclusive processes.

In this proposed experiment we plan to use a PbWO$_4$ calorimeter with a resolution more than twice better than the PbF$_2$ calorimeter used in E00-110. While the missing mass resolution will be slightly worse at some high beam energy, low $x_B$ kinematics, the better energy resolution of the crystals will largely compensate for it, and the missing mass resolution in this experiment will be significantly better than ever before. Fig.~\ref{fig:mm2} (right) shows the missing mass resolution for PbF$_2$ and PbWO$_4$ for a kinematic setting similar to the one measured in Hall A. Tab.~\ref{tab:DVCS-Kin} shows the missing mass resolution projected for each of the settings using the proposed PbWO$_4$ calorimeter. 

\subsection{Systematics uncertainties}

\begin{table}[t]
\begin{center}
 \begin{tabular}{||l|c|c||} 
 \hline
  \hspace*{1.5cm}Source
  &  pt-to-pt    &  scale  \\
                                      &  (\%)        &  (\%)   \\
 \hline 
  Acceptance                   &   0.4        &   1.0   \\
  Electron/positron PID                 &   $<$0.1     &   $<$0.1    \\
  Efficiency
       &   0.5        &   1.0  \\
  Electron/positron tracking efficiency &   0.1        &   0.5    \\
  Charge                       &   0.5        &   2.0    \\
  Target thickness             &   0.2        &   0.5    \\
  Kinematics                   &   0.4        &  $<$0.1    \\
  Exclusivity                  &     1.0        &   2.0 \\
  $\pi^0$ subtraction &  0.5          &   1.0\\
  Radiative corrections        & 1.2 & 2.0\\
  \hline 
  Total                       &   1.8--1.9       &   3.8--3.9    \\
 \hline
 \end{tabular}
\end{center}
\caption{\baselineskip 13pt Estimated systematic uncertainties for the proposed experiment based on previous Hall C experiments.}
\label{tab:sys} 
\end{table}

The HMS is a very well understood magnetic spectrometer which will be used here with modest requirements (beyond the momentum), defining the ($x_B,Q^2$) kinematics well. Tab.~\ref{tab:sys} shows the estimated systematic uncertainties for the proposed experiment based on previous experience from Hall C equipment and Hall A experiments.

\section{Proposed kinematics and projections}
\label{sec:Projections}
Table~\ref{tab:DVCS-Kin} details the kinematics and beam time used in the projection. $Q^2$ scans at 4 different values of $x_B$ were chosen in kinematics with already approved electron data~\cite{E12-13-010}. The positron beam current assumed is 5~$\mu A$ (unpolarized beam) and is currently the limiting factor driving the beam time needs. %
Considering a 1~$mA$ initial electron beam, 5~$\mu A$ of positrons corresponds to the maximum positron to electron ratio produced by 123 MeV electrons~\cite{Cardman:2018svy}. This also corresponds to the lowest polarization transfer to the positrons, which will be considered unpolarized in our projections. %
Beam time in Table~\ref{tab:DVCS-Kin} is calculated in order collect positron data corresponding to $\sim$25\% of the approved electron data.

\begin{table*}[t]
\begin{center}
\hspace*{-0.4cm}\begin{tabular}{|c|c|c|c|c|c|c|c|c|c|c|c|c|c|c|c|c|c|}  
\hline
 \hline
$x_\text{Bj}$ &\multicolumn{4}{c|}{0.2}&\multicolumn{6}{c|}{0.36}&\multicolumn{3}{c|}{0.5}&\multicolumn{4}{c|}{0.6}\\ \hline
                \hline
                $Q^2\,\text{(GeV)}^2$ &\multicolumn{3}{c|}{2.0}&3.0&\multicolumn{3}{c|}{3.0}&\multicolumn{2}{c|}{4.0}&5.5&\multicolumn{2}{c|}{3.4}&4.8&\multicolumn{3}{c|}{5.1}&6.0\\\hline
                $E_b\ \text{(GeV)}$ & 6.6&8.8&\multicolumn{2}{c|}{11}&6.6&8.8&11&8.8&\multicolumn{2}{c|}{11}&8.8&\multicolumn{2}{c|}{11}&6.6&8.8&\multicolumn{2}{c|}{11}\\
                \hline
                $k'\ \text{(GeV)}$ &1.3&3.5&5.7&3.0&2.2&4.4&6.6&2.9&5.1&2.9&5.2&7.4&5.9&2.1&4.3&6.5&5.7\\\hline
                $\theta_\text{Calo}\,\text{(deg)}$ &6.3&9.2&10.6&6.3&11.7&14.7&16.2&10.3&12.4&7.9&20.2&21.7&16.6&13.8&17.8&19.8&17.2\\\hline
                $D_\text{Calo}$ (m) & 6&\multicolumn{2}{c|}{4}&6&\multicolumn{3}{c|}{3}&4&3&4&\multicolumn{7}{c|}{3}\\\hline
                $\sigma_{M_X^2}$(GeV$^2$) &\multicolumn{3}{c|}{0.17}&0.22&\multicolumn{2}{c|}{0.13}&0.12&\multicolumn{2}{c|}{0.15}&0.19&\multicolumn{2}{c|}{0.09}&0.11&\multicolumn{4}{c|}{0.09}\\\hline
                $I_\text{beam}$ ($\mu$A)&\multicolumn{17}{c|}{5}\\\hline
            Days &    1	&1&	3&	1&	2&	3&	2&	3&	4&	13&	4&	3&	7&	7&	2	&7&	14\\\hline

\hline
\end{tabular}
\end{center}
\caption{\baselineskip 13pt
DVCS kinematics with positrons in Hall C.  The incident and scattered beam energies are $k$ and $k'$, respectively.  The calorimeter
is centered at the angle $\theta_\text{Calo}$, which is set equal to the nominal virtual-photon direction.  The front face of the calorimeter
is at a distance $D_\text{Calo}$ from the center of the target,  and is adjusted to optimize multiple parameters:  First to maximize acceptance,
second to ensure sufficient separation of the two clusters from symmetric $\pi^0\rightarrow \gamma\gamma$ decays, and third to ensure
that the edge of the calorimeter is never at an angle less than $3.2^\circ$ from the beam line. The maximum expected positron beam current (5~$\mu$A) will be used for all kinematics settings. The total amount of beam time needed is 77 days.}
\label{tab:DVCS-Kin}
\end{table*}%

\begin{figure}[htb]
  \centering\includegraphics[width=\linewidth]{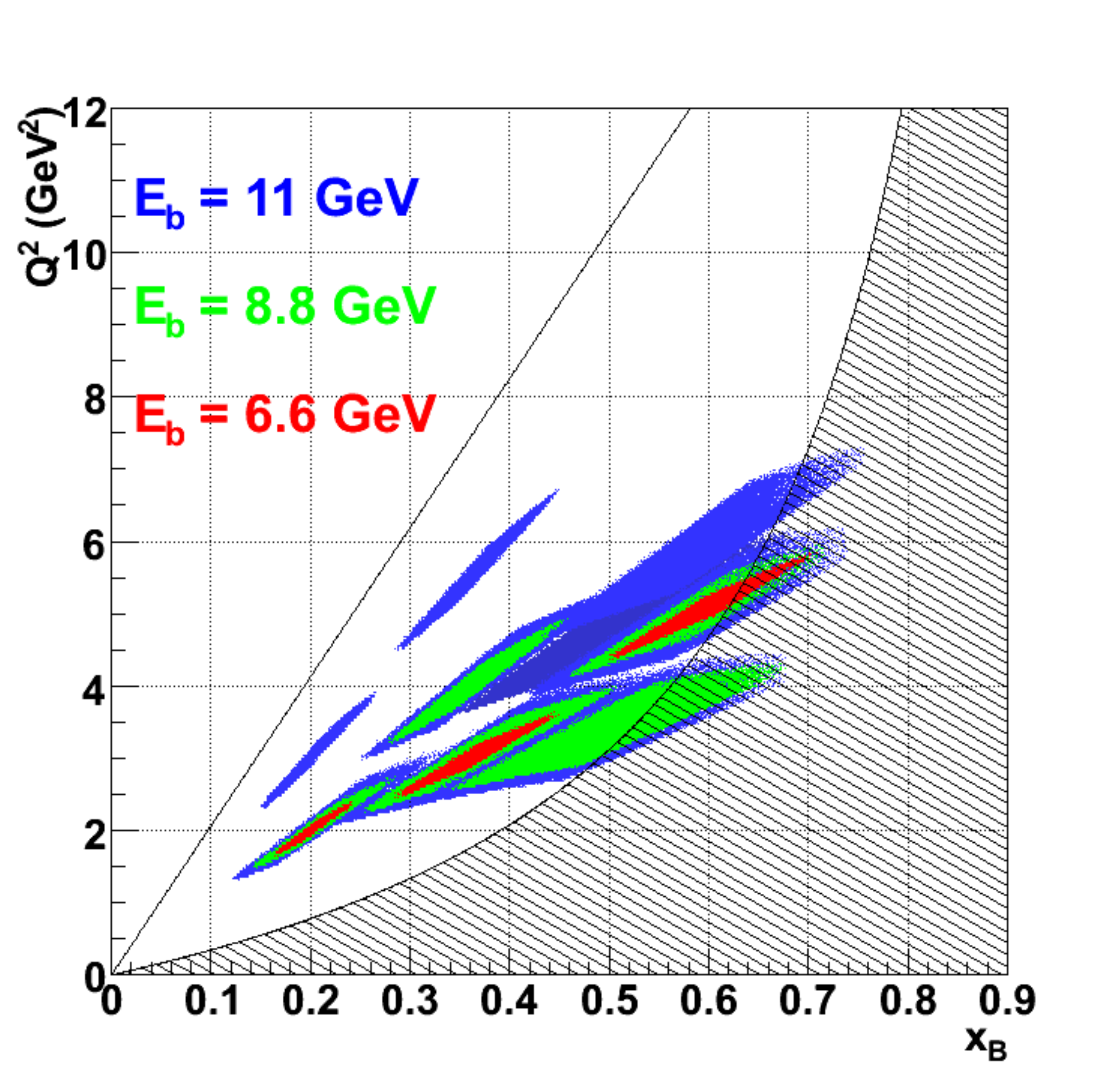}
\caption{
Display of different kinematic setting proposed. The $Q^2-x_B$ settings correspond to the ones approved in experiment E12-13-010, which will measure DVCS cross sections using an electron beam. Shaded areas show the resonance region $W<2$~GeV and the line $Q^2=(2M_pE_b)x_B$ limits the physical region for a maximum beam energy $E_b=11$~GeV.}
\label{fig:kin}
\end{figure}

The different kinematics settings are represented in Fig.~\ref{fig:kin} in the $Q^2$--$x_B$ plane. The area below the straight line $Q^2=(2M_pE_b)x_B$ corresponds to the physical region for a maximum beam energy $E_b=11$~GeV. Also plotted is the resonance region $W<2$~GeV.

We have performed detailed Monte Carlo simulation of the experimental setup and evaluated counting rates for each of the settings. In order to do this, we have used a recent global fit of world data with LO sea evolution by D.~M\"uller and K.~Kumeri\v cki~\cite{DMweb}. This fit reproduces the magnitude of the DVCS cross section measured in Hall A at $x_B=0.36$ and is available up to values of $x_B\le 0.5$. For our high $x_B$ settings we used a GPD parametrization by P.~Kroll, H.~Moutarde and F.~Sabati\'e~\cite{Kroll:2012sm} fitted to Deeply Virtual Meson Production data, together with a code to compute DVCS cross sections, provided by H.~Moutarde~\cite{HerveTGV,TGGuichon}. Notice that for DVCS, counting rates and statistical uncertainties will be driven {\em at first order} by the Bethe-Heitler (BH) cross section, which is well-known.

\begin{figure*}[htb]
\includegraphics[width=1\linewidth]{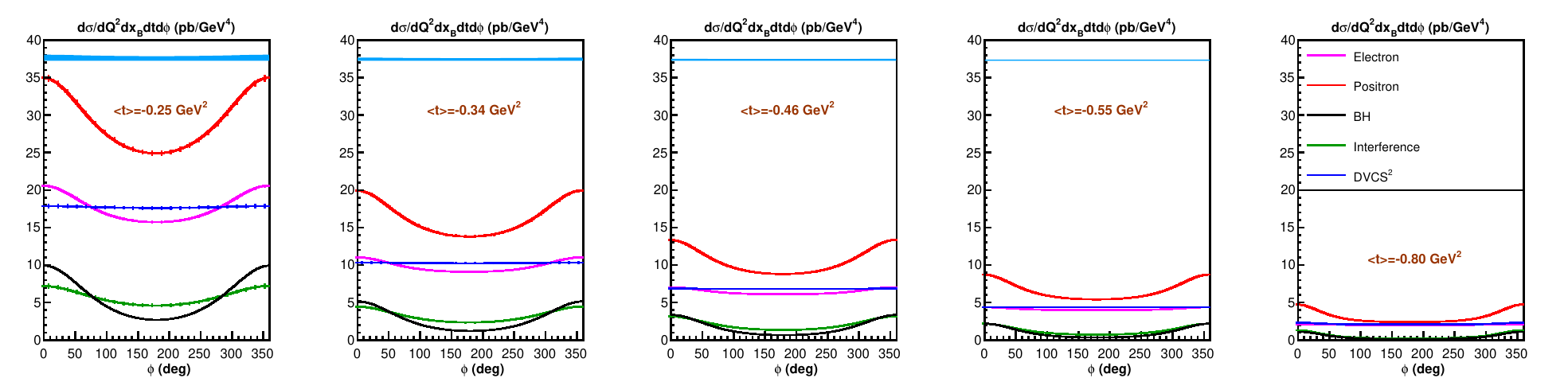}\\
\includegraphics[width=1\linewidth]{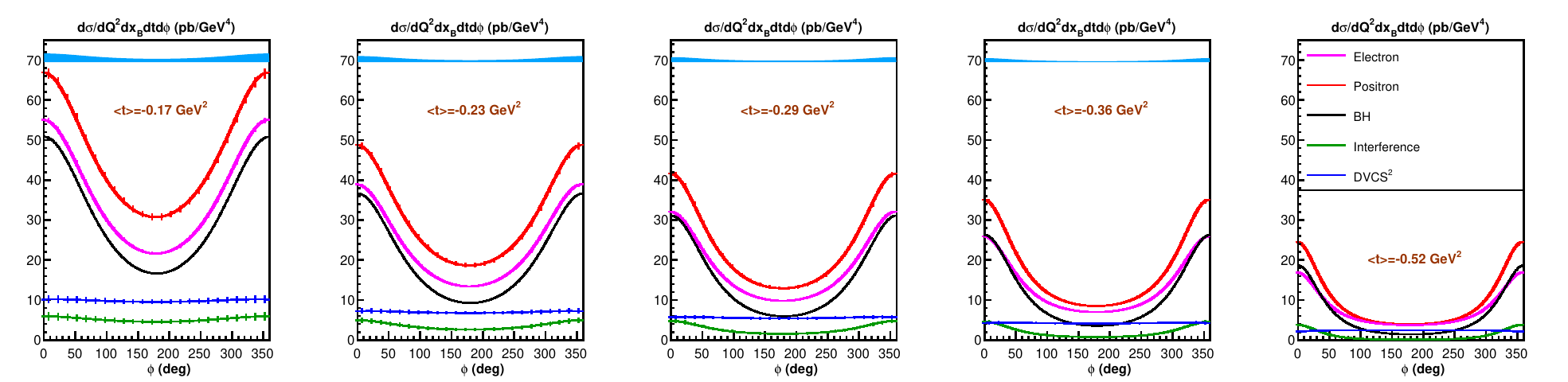}\\
\includegraphics[width=1\linewidth]{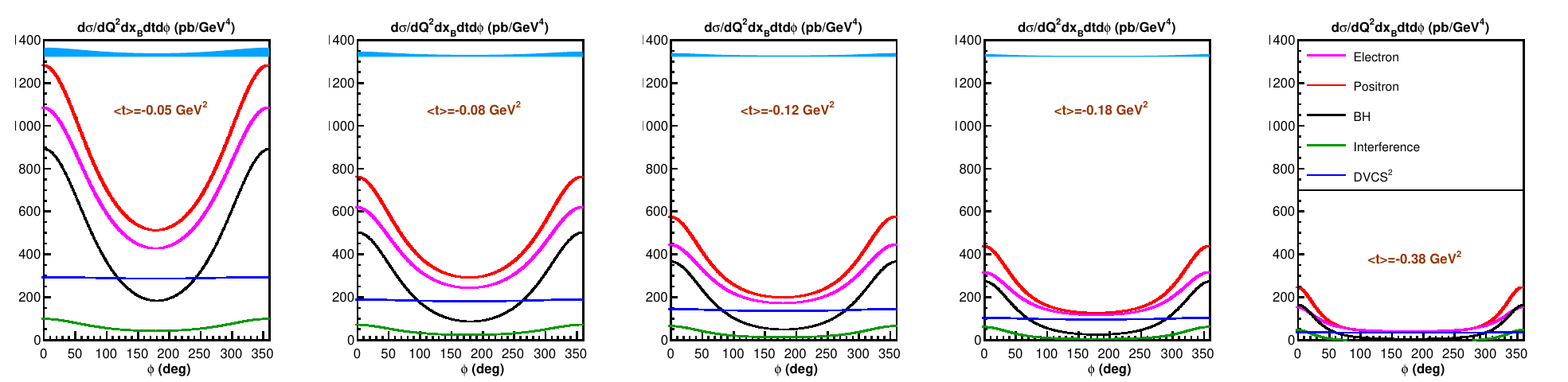}
\caption{\baselineskip 13 pt
Experimental projections for 3 different of the settings proposed: $x_B=0.2$, $Q^2=2.0$~GeV$^2$ (top),  $x_B=0.36$, $Q^2=4.0$~GeV$^2$ (middle) and $x_B=0.5$, $Q^2=3.4$~GeV$^2$ (bottom). Red points show the projected positron cross sections with statistical uncertainties. Electron cross sections that will be measured in experiment E12-13-010 are shown in magenta. The combination of $e^-$ and $e^+$ cross sections allow the separation of the DVCS$^2$ contribution (blue) and the DVCS-BH interference (green). For reference, the BH cross section is displayed in black. Systematic uncertainties are shown by the cyan band.}
\label{fig:ros}
\end{figure*}

Fig.~\ref{fig:ros} shows the projected results for 3 selected settings at different values of $x_B=0.2, 0.36, 0.5$. Statistical uncertainties are shown by error bars and systematic uncertainties are represented by the cyan bands.

The DVCS$^2$ term (which is $\phi$ independent at leading twist) can be very cleanly separated from the BH-DVCS interference contribution, and this without any assumption regarding the leading-twist dominance. The $Q^2-$de\-pen\-dence of each term will be measured (cf. Tab.~\ref{tab:DVCS-Kin}) and its dependence compared to the asymptotic prediction of QCD. The extremely high statistical and systematic precision of the results illustrated in Fig.~\ref{fig:ros} will be crucial to disentangle higher order effects (higher twist or next-to-leading order contributions) as shown by recent results~\cite{Defurne:2017paw}.

\section{Constraints on Compton Form Factors}
In order to quantify the impact of the proposed experiment on the extraction of the nucleon Compton Form Factors, we have simulated the extraction of the proton CFFs by using only approved electron cross-section measurements (both helicity-dependent and helicity-independent) from upcoming experiment E12-13-010 and with the addition of the positron measurements proposed herein. 
Measurements with an unpolarized target as proposed herein have little sensitivity to GPDs $E$ and $\widetilde E$. Therefore, only the CFFs corresponding to $H$ and $\widetilde H$ have been fitted. Prospects of measurements with polarized targets would be, of course, extremely exciting and complementary to these. Most importantly, as mentioned before, kinematics corrections of $\mathcal O(t/Q^2)$ and $\mathcal O(M^2/Q^2)$ cannot be neglected in JLab kinematics. Therefore, all CFFs $\mathbb H_{++}$, $\mathbb H_{0+}$, $\mathbb H_{-+}$, $\widetilde{\mathbb H}_{++}$, $\widetilde{\mathbb H}_{0+}$ and $\widetilde{\mathbb H}_{-+}$ have been fitted.

\begin{figure*}
\centering\includegraphics[angle=90, width=0.7\linewidth]{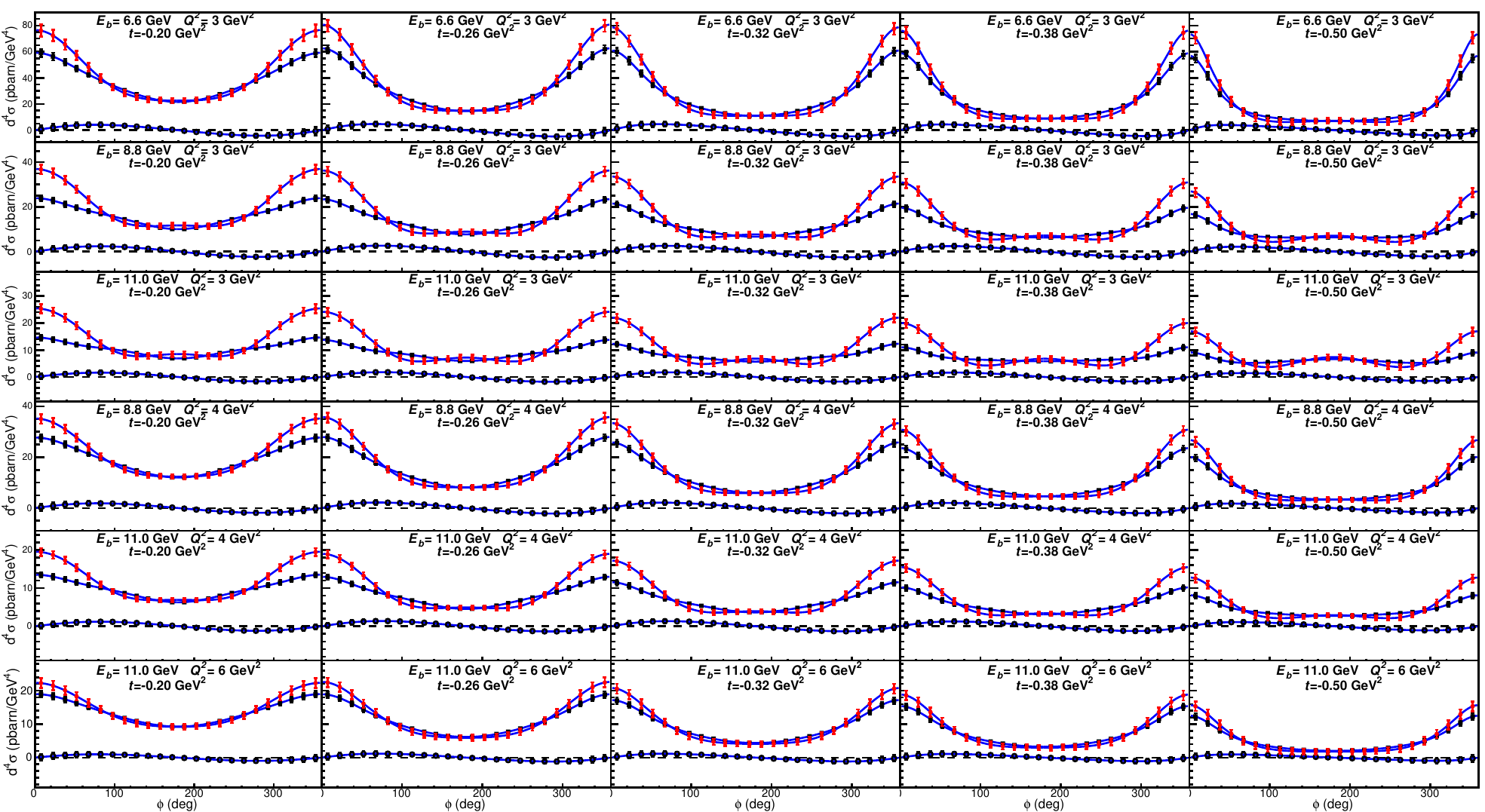}
\caption{Fits of data with $x_B=0.36$. Each row shows five kinematic settings in $Q^2$ at constant $x_B$ and $t$. Each column corresponds to 5 different bins in t. The blue lines are simultaneous fits of the (approved) electron data, both the helicity-dependent (black squares) and helicity-independent (black circles) cross sections, and the (proposed) positron data (red points). Horizontal dashed lines in each panel indicate the origin of the vertical axis: $d^4\sigma=0$.}
\label{fig:xsecfit}
\end{figure*}

First of all, the DVCS cross sections measured in Hall A with a 6 GeV beam~\cite{Defurne:2015kxq,Defurne:2017paw} were fitted in order to extract some realistic values of the CFFs. These values were then used to calculate projected cross sections at the kinematics of Tab.~\ref{tab:DVCS-Kin}. The CFFs are assumed constant in $t$ for this exercise and
equal to the average value of those extracted from 6 GeV data. The projected electron and positron cross sections are then fitted. In doing this, the statistical and systematic uncertainties of the measurements were added quadratically. Fig.~\ref{fig:xsecfit} shows the results for kinematics with $x_B=0.36$. Each line shows the five
kinematic settings in {$Q^2$, $E_b$} at constant $x_B$ and $t$, which are fitted simultaneously neglecting the logarithmic $Q^2$-dependence of CFFs in the range of $\sim$3--6 GeV$^2$. In addition to the five independent terms on the azimuthal angle ($\sim 1$, $\sim\cos{\phi}$, $\sim\cos{2\phi}$, $\sim\sin{\phi}$ and $\sim\sin{2\phi}$), three different beam energies are fitted simultaneously.
Each column in Fig.~\ref{fig:xsecfit} shows each of the 5 bins in $t$ where the data were binned. The blue lines correspond to the fits of both the (approved) electron data (helicity-dependent and helicity-independent) and the positron (proposed) data (only helicity-independent). Notice that the NPS calorimeter acceptance will allow a full coverage in $\phi$ for the bins in $t$ presented.


\begin{figure*}[!hbt]
\includegraphics[width=0.66\linewidth]{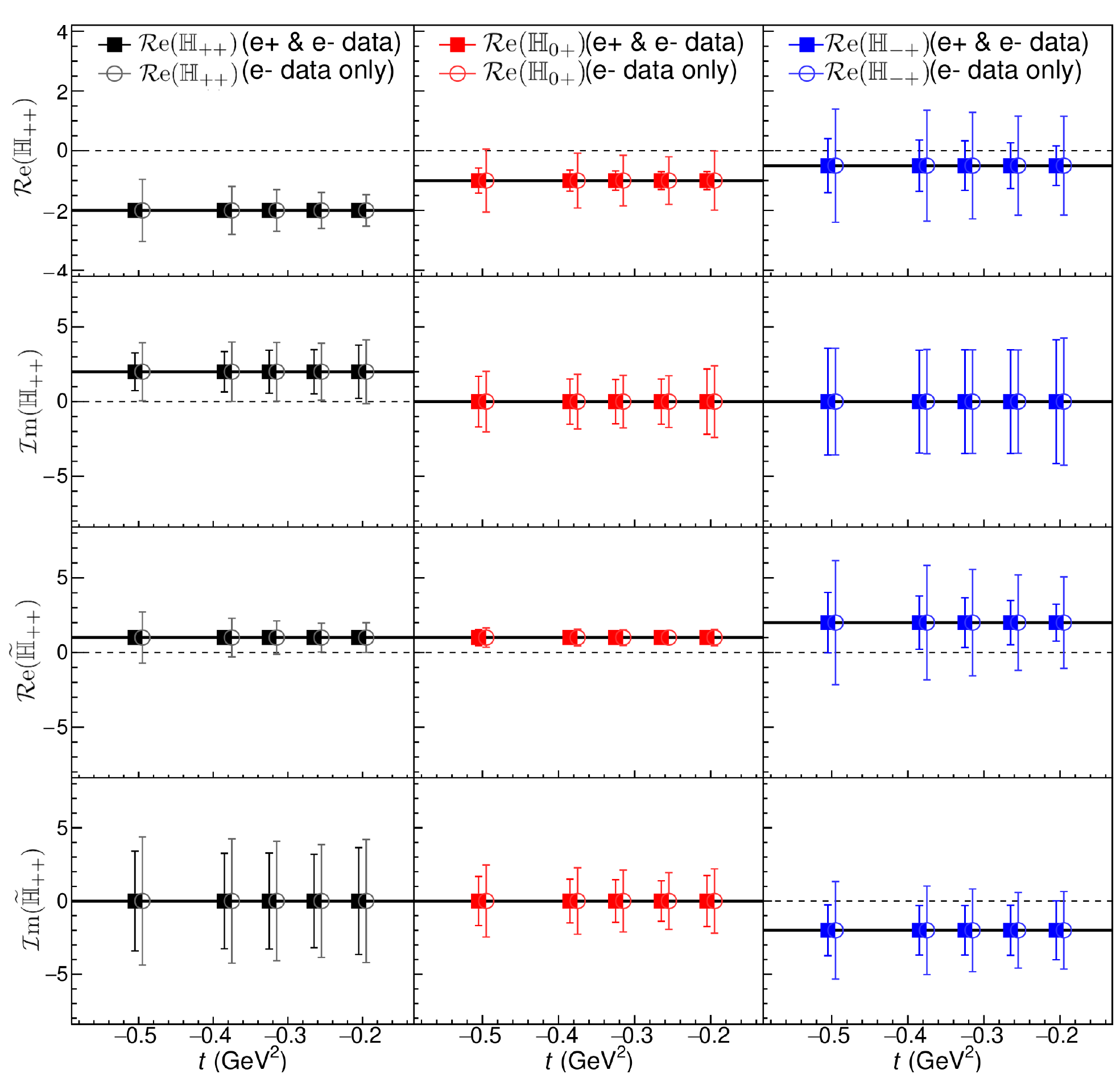}
\includegraphics[width=0.33\linewidth]{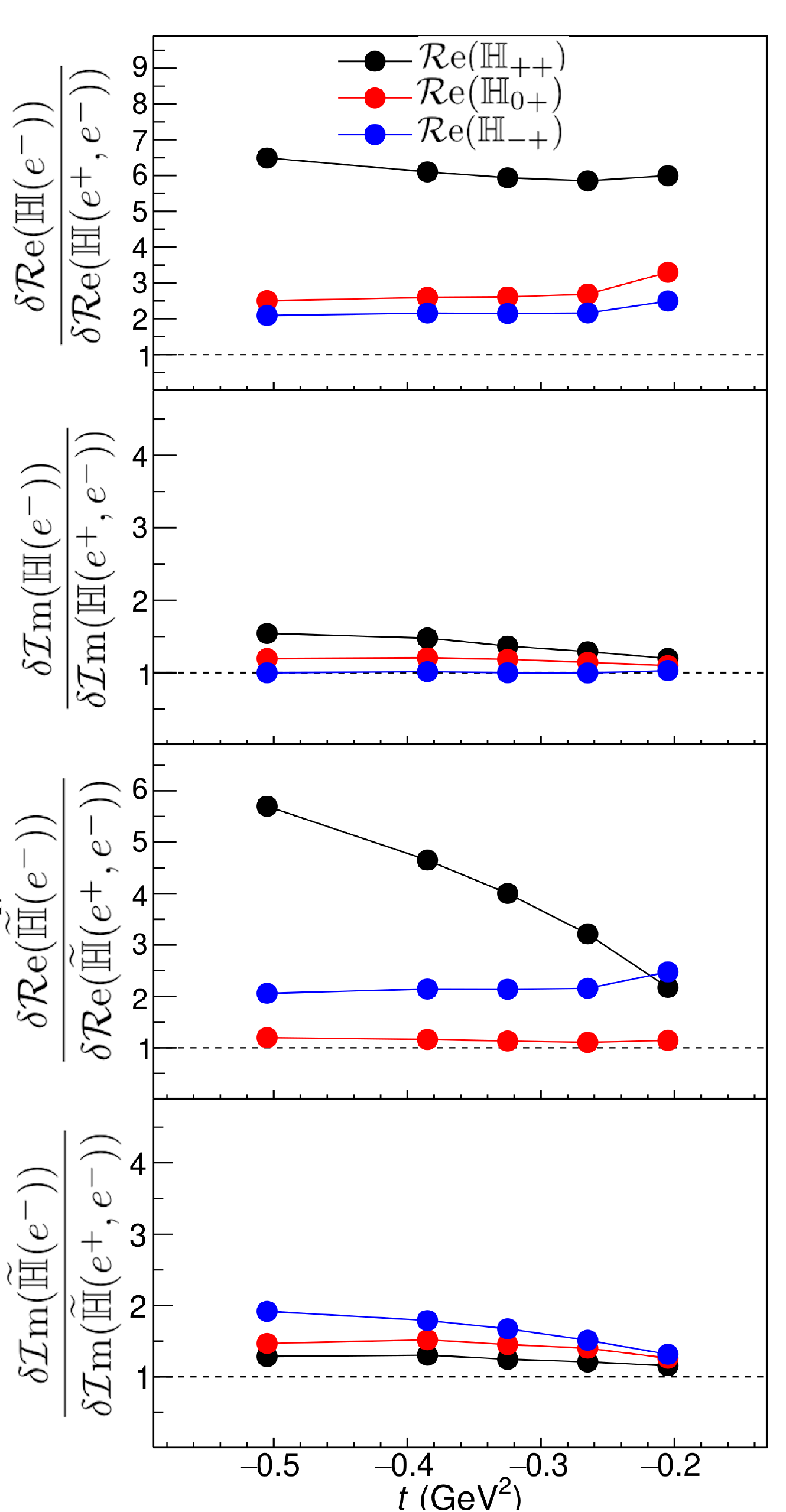}
\caption{CFFs extracted from the fits in Fig.~\ref{fig:xsecfit}. Left: the first column in the left shows the results of the helicity-conserving CFFs when both positron and electron data are used in the fit (black), and when only the electron approved data is used (grey). The second and third columns show the same information for the helicity-flip CFFs. The solid horizontal lines indicate the input values used to generate the cross-section data. Right: ratio of the uncertainties between the fit using both electron and positron data and the one using only electron data.}
\label{fig:fits}
\end{figure*}

Results of the CFFs extracted from the fits are shown in Fig.~\ref{fig:fits}. The first column in the left shows the results of the helicity-conserving CFFs when both positron and electron data are used in the fit, and when only the electron approved data are used. The second and third columns show the same information for the helicity-flip CFFs. The solid horizontal lines in each panel indicate the input values used to generate the cross-section data, which are then accurately extracted by the fit. The ratio of the uncertainties between the fit using both electron and positron data and the one using only electron data is shown in the last column on the right. One can see the significant improvement of positron data: a factor of 6 for $\mathcal Re(\mathbb H_{++})$ and an average factor of 4 for $\mathcal Re(\widetilde{\mathbb H}_{++})$. There is also a factor $\sim$2 improvement in the real part of most helicity-flip CFFs. The imaginary part of CFFs are not impacted by these positron data -- this is expected as no helicity-dependent positron cross sections are used in the fits.

\begin{figure}[!hbt]
\includegraphics[width=\linewidth]{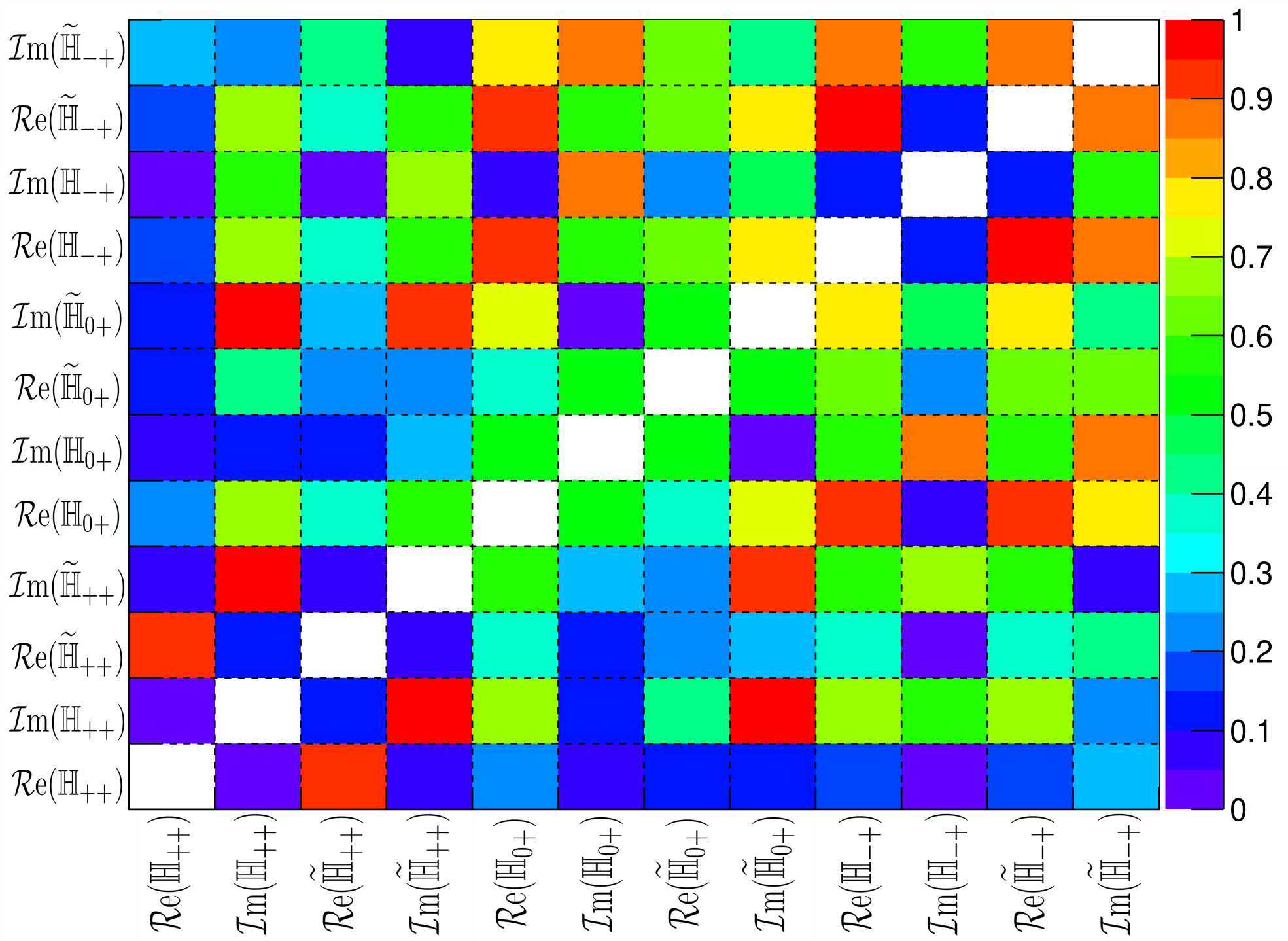}\\
\includegraphics[width=\linewidth]{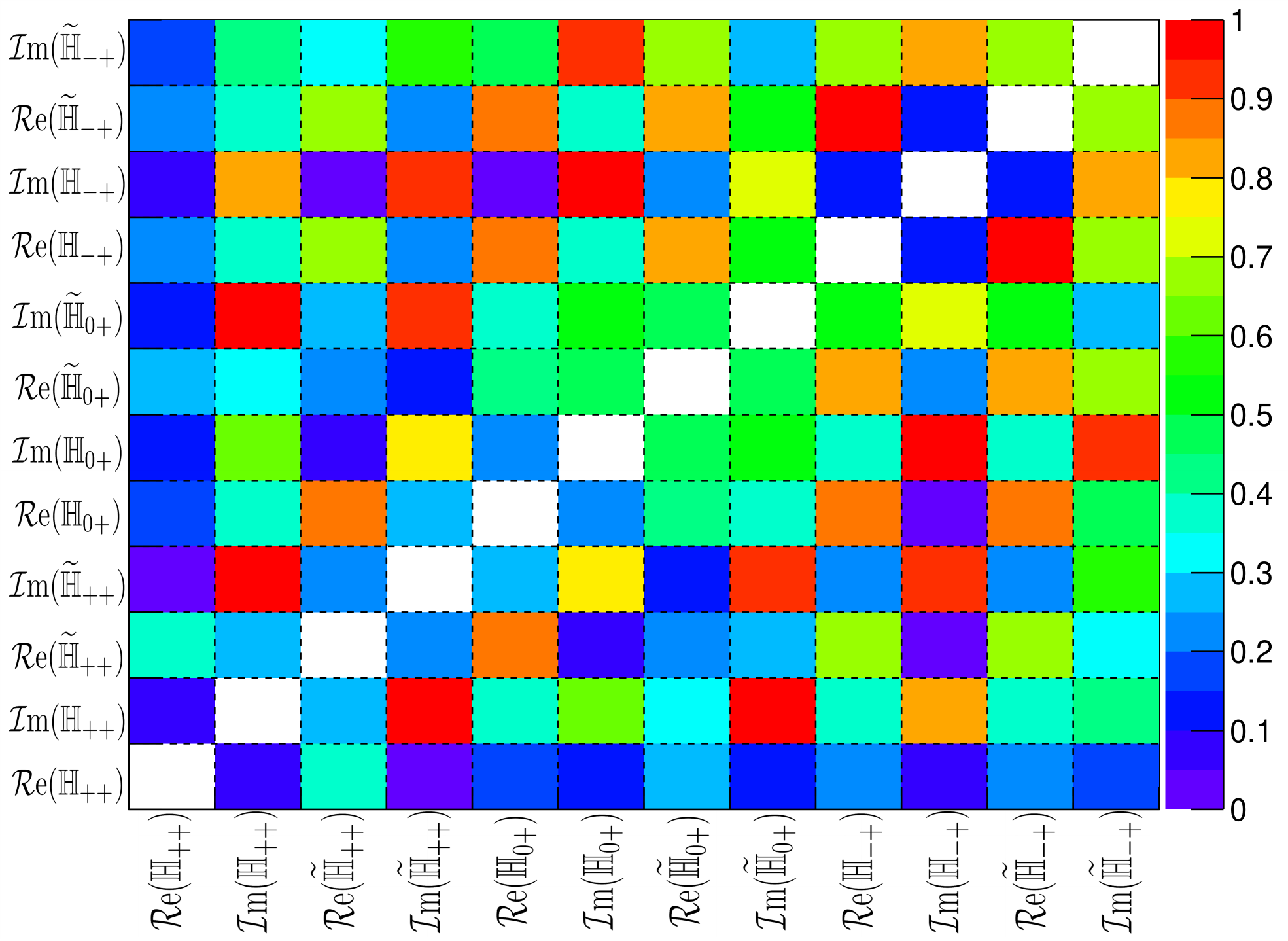}
\caption{Magnitude of the correlation coefficients between the different CFFs extracted from the fit of DVCS electron data (top) and from the combined fit of DVCS electron and positron data (bottom). Plots correspond to bin $x_B=0.36$ and $t=-0.26$~GeV$^2$. The correlation between $\mathcal Re(\mathbb H_{+,+})$ and $\mathcal Re(\widetilde{\mathbb H}_{+,+})$ goes from $-94\%$ without positrons to $-39\%$ when electron and positrons are combined.}
\label{fig:corr}
\end{figure}

In addition to reducing the uncertainties of the fitted CFFs, positron data also improves the correlation of the extracted parameters. Fig.~\ref{fig:corr} shows the correlation coefficient between the different pairs of CFFs as extracted from the electron data alone (left) and with the addition of positron data (right). The correlation coefficient for each pair of extracted CFFs $(\mathbb F_i, \mathbb F_j)$ is defined as $\rho_{i,j}=\text{cov}[\mathbb F_i, \mathbb F_j]/(\sigma_i\sigma_j)$. It varies from -1 to 1. and Fig.~\ref{fig:corr} reports its absolute value. One can notice, in particular, that while the helicity-conserving real parts of $\mathbb H_{+,+}$ and $\widetilde{\mathbb H}_{+,+}$ are very correlated in the case of a fit with electron data only, the correlation is significantly reduced when positron data are included. The improvement varies from $-98\%$ to $-54\%$ at the highest value of $|t|$ and from $-70\%$ to $-24\%$ at the lowest $|t|$.

\section{Summary}
We propose to measure the cross section of the DVCS reaction accurately using positrons in the wide range of kinematics allowed by a set of beam energies up to 11 GeV. We will exploit the beam charge dependence of the cross section to separate the contribution of the BH-DVCS interference and the DVCS$^2$ terms.

The $Q^2-$dependence of each individual term will be measured and compared to the predictions of the handbag mechanism. This will provide a quantitative estimate of higher-twist effects to the GPD formalism in JLab kinematics.

The combination of measurements with electrons and positrons allow to much better constrain the Compton Form Factors measurements and reduce significantly the correlations in the extracted values.

We plan to use Hall C High-Momentum Spectrometer, combined with a high resolution PbWO$_4$ electromagnetic calorimeter.

In order to complete this full mapping of the DVCS cross section with positrons over a wide range of kinematics, we require 77 days of (unpolarized) positron beam (I$>5 \mu$A).

\section{Acknowledgements}
This material is based upon work supported by the U.S. Department of Energy, Office of Science, Office of Nuclear Physics under contract DE-AC05-06OR23177.

\bibliographystyle{unsrt}
\bibliography{Compton2014}
\end{document}